\documentclass[prd,twocolumn,showpacs,floatfix,
superscriptaddress,preprintnumbers , letterpaper ]{revtex4}
\usepackage{graphicx}
\usepackage{epsfig}
\usepackage{bm}
\usepackage{amsfonts}

\begin{document}

\def\be{\begin{equation}}
\def\ee{\end{equation}}
\def\bea{\begin{eqnarray}}
\def\eea{\end{eqnarray}}

\title{A new approach to modified gravity models }
\author{Sayan K. Chakrabarti\footnote{Present address: CENTRA, Departamento
de Fisica,\\
Instituto Superior Tecnico - IST\\
Lisbon, Portugal\\
email: sayan.chakrabarti@ist.utl.pt}}
\affiliation{Theory Division, Saha Institute of Nuclear Physics, 1/AF
Bidhannagar, Kolkata 700064, India}
\author{Emmanuel N. Saridakis\footnote{Present Address: College of
Mathematics
and Physics,\\ Chongqing University of Posts and
Telecommunications\\ 
Chongqing 400065, P.R. China, email: msaridak@phys.uoa.gr}}
\affiliation{Department of Physics, University of Athens, GR-15771
Athens, Greece}
\author{Anjan A. Sen \footnote{anjan.ctp@jmi.ac.in}}
\affiliation{Center For Theoretical Physics,
Jamia Millia Islamia, New Delhi 110025, India}

\begin{abstract}
We investigate $f(R)$-gravity models performing the ADM-slicing of
standard General Relativity. We extract the static,
spherically-symmetric vacuum solutions in the general case, which
correspond to either Schwarzschild de-Sitter or Schwarzschild
anti-de-Sitter ones. Additionally, we study the cosmological
evolution of a homogeneous and isotropic universe, which is
governed by an algebraic and not a differential equation. We show
that the universe admits solutions corresponding to  acceleration
at late cosmological epochs, without the need of fine-tuning the
model-parameters or the initial conditions.
\end{abstract}


\maketitle

\section{Introduction}

Recently, there has been  a tremendous thrust in research
activities in cosmology, due to the wealth of data from various
experiments that are already available and more that are
anticipated in the near future. Amongst others, the discovery of
the late time acceleration of the universe has been particularly
fascinating for cosmologists, particle physicists and string
theorists alike. In particular, the significant improvement of the
data-statistics (from 50 to more than 300 Supernova(SnIa))
\cite{snIa} has made the aforementioned result indisputable.
Furthermore, complementary probes like Baryon Acoustic Oscillation
\cite{bao} and Cosmic Microwave Background Radiation measurements,
\cite{wmap} offer additional evidences.

Although the simplest candidate of the acceleration mechanism is
the cosmological constant, one can construct various ``field''
models in order to incorporate this mysterious ``dark energy'' In
particular, one can use a canonical scalar field (quintessence)
\cite{quintess}, a scalar field with non-standard kinetic
term (k-essence) \cite{kessence} or with a negative sign of
the kinetic term (phantom) \cite{phant}, the combination of
quintessence and phantom in a unified model named quintom
\cite{quintom}, scalar fields non-minimally coupled to gravity
\cite{nonminimal} or simple barotropic fluids with specific
pressure-form such as Chaplygin gas \cite{Chalpygin} (for nice
reviews on dark energy models, see \cite{rev}).

Instead of using field dark energy constructions, an alternative
approach is to modify gravity itself. The Dvali-Gabbadadge-Poratti
(DGP) \cite{dgp}, the Cardassian \cite{cardassian} and the
Shtanov-Sahni \cite{shtanov} models follow this direction, lying
in particular in the higher-dimensional scenario sub-class.
However, instead of using extra dimensions, one can insert
higher-order curvature invariants in the usual Einstein-Hilbert
action, with the simple consideration the so-called $f(R)$-gravity
models \cite{fr}, that is the addition of Ricci-scalar functions.
Such models wish to alleviate the non-renormalizability of
gravity, and acquire theoretical justification from low-energy
string theory. Alternatively, since higher-order terms can be
related to non-minimally coupled scalar degrees of freedom, these
models are equivalent to scalar-tensor constructions. Concerning
their cosmological implications, they can describe both inflation
as well as the late-time acceleration. However, most of
$f(R)$-gravity models do not manage to pass the observational and
theoretical tests (solar system, neutron stars and binary pulsar
constraints), giving also rise to an unusual matter dominated
epoch and leading to significant fine-tunings \cite{ft}. Moreover,
even if improved versions (like the ``Chameleon mechanism''
\cite{ch}) manage to satisfy the above constraints, one can still
face problems in the strong gravity regime, due to the curvature
singularity that is inevitable at the nonlinear level \cite{nl}.
 But, it was lately shown that $f(R)$ gravity may contain all four known
types of future
 singularities
namely the Big Rip, type II, III or IV \cite{nonsing} and it was shown that
one of the ways to 
avoid future singularity via unification of inflation with dark energy (see
\cite{revodin} for
 a detailed review) is the introduction of $R^2$ term. 
Let us also mention in this context here that recently in \cite{cenosz}
some realistic $f(R)$
 models were proposed which successfully passes the local tests and
fulfills the cosmological
 bounds (see also
\cite{reviewsdosn} for a review).

Although higher time-derivatives can be beneficial in making
gravity renormalizable, they also lead to ghosts. However, the
recently developed Ho\v{r}ava gravity wishes to act as a
power-counting renormalizable, Ultra-Violet (UV) complete theory
of gravity  \cite{horava1}, without possessing the full
diffeomorphism invariance of General Relativity but only a subset
that is manifest in the Arnowitt, Deser and Misner  (ADM) slicing.
There has been a large amount of effort in examining and extending
its properties, as well as exploring its cosmological implications
\cite{HLaspects}.

Motivated by these, in the present work  we are interested in
investigating  $f(R)$ gravity models in purely metric gravity
 performing the ADM slicing of
standard General Relativity, that is its (3+1)-decomposition based
on the Hamiltonian formulation \cite{adm}. In particular, we wish
to extract  the static, spherically-symmetric vacuum solution for
general $f(R)$-models under ADM decomposition, and study the
homogeneous and isotropic cosmological solutions which present
late-time acceleration. The paper is organized as  follows: In
section \ref{ADM} we present a brief introduction on ADM
(3+1)-decomposition and we write the gravitational action for
$f(R)$-gravity models. In section \ref{Static} we derive the
static, spherically-symmetric vacuum solutions for general $f(R)$
models under ADM slicing. In section  \ref{cosmo} we apply this
approach to cosmological frameworks, investigating universe
evolutions that experience late-time acceleration. Finally,
section \ref{Conclusions} is devoted to the summary of the
obtained results.

\section{ADM (3+1)-decomposition and $f(R)$ gravity models}
\label{ADM}

We start by considering the ADM decomposition of the four
dimensional metric \cite{adm}. Any arbitrary global manifold
${\mathcal{M}}$ can be foliated in a family of hypersurfaces $\Sigma$ with
constant $t$ denoted by $\Sigma_t$. 
Assuming the topoplogy of the spacetime to be of the form $\Sigma\times
{\mathbf{R}}$, the metric can be split into spatial and time component.  
\begin{equation}
\label{metric0}
 ds ^{2} = -N^{2}
dt^{2} + g_{ij}(dx^{i} + N^{i} dt)(dx^{j}+N^{j}dt).
\end{equation}
Here $N$ and $N^{i}$ are the lapse function and  shift vectors
respectively, while $g_{ij}$ is the induced metric on the
3-dimensional space-like hypersurface for a fixed time.  Indices
of all projected tensors can be lowered (raised) by $g_{ij}$
($g^{ij}$).

For the space-like hypersurface with fixed time, extrinsic
curvature is defined as
 \begin{equation}
  K_{ij} =
{1\over{2N}}(\dot{g_{ij}} - \nabla_{i}N_{j}-\nabla_{j}N_{i}),
\end{equation}
where {\it dot} represents derivative with respect
to time and $i,j= 1,2,3$. Under such decomposition,
\begin{equation}
 R^{(4)} = R^{(3)} + K_{ij}K^{ij} - K^{2} +
2\nabla_{\mu}\left(n^{\mu}\nabla_{\nu}n^{\nu} -
n^{\nu}\nabla_{\nu}n^{\mu}\right),
  \end{equation}
where  $R^{(4)}$ and $R^{(3)}$ are the four and  three-dimensional
Ricci scalars respectively and $n^{\mu}$ is a unit vector perpendicular to
the three dimensional hypersurface defined by $t=\rm{constant}$ and
$K=g^{ij}K_{ij}$ is the trace of the extrinsic curvature.

In standard Einstein gravity, the last term in r.h.s of equation (3) is a
total derivative term and does not contribute to the equation of motion,
hence the Einstein-Hilbert action can
now be written as
\begin{equation}
{\cal S} = \int {1\over{16\pi G}} dt d^{3}x
N\sqrt{g^{(3)}} (R^{(3)} + K_{ij}K^{ij} - K^{2}) + {\cal S}_{m},
\end{equation}
 where $g^{(3)}$ is the determinant of the
three-dimensional space-like hypersurface, and ${\cal S}_{m}$
accounts for the matter content of the universe. We should mention that
this does not happen for modified action which is proportional to
$f(R^{(4)})$ where the contribution from the last term of equation (3) can
not be in general written as a total derivative term. 

With this, our modified gravity action is given by,
\begin{eqnarray}
S &=& \frac{1}{16\pi G}\int dt d^3x N\sqrt{g^{(3)}}\Big[R^{4}
+ F\Big(R^{(3)}+K_{ij}K^{ij}\nonumber\\&&-K^2\Big)\Big]+S_m,
\label{actionfinal}
\end{eqnarray}

where $R^{(4)}$ is given by equation (3) and we have ignored the term
involving derivatives of $n^{\mu}$ in the modified part of the action. This
action clearly breaks the full diffeomorphism in the 4-d space-time, but it
preserves the foliation preserving diffeomorphism in the 3-d space-like
hypersurface. The structure of the action is same as that proposed in the
Ho\v{r}ava Gravity \cite{horava1}.

Our main motivation for such modification is to explain the late time
acceleration of the universe but before studying cosmology in this setup,
we discuss the vacuum spherically symmetric static solutions for the action
(5).

\section{Static spherically symmetric solutions}
\label{Static}

We are looking for the static, spherically-symmetric vacuum
solutions of the aforementioned general $f(R)$ gravity models,
under ADM decomposition. In this case the metric (\ref{metric0})
writes:
 \begin{equation}
ds^2=-N^2(r)dt^2+\frac{1}{h(r)}dr^2+ r^2d\Omega^2,
\end{equation}
that is $K_{ij}=0$ and
$R^{(3)}=-\frac{2}{r^2}[rh^{\prime}(r)+h(r)-1]$. Setting
$h(r)-1\equiv X(r)$, we obtain
$rh^{\prime}(r)+h(r)-1=rX^{\prime}(r)+X(r)=[X(r)r]^{\prime}$. Thus,
defining $\mathcal{H}(r)$ as
 \begin{equation}
R^{(3)}=-\frac{2}{r^2}[X(r)r]^{\prime}\equiv \mathcal{H}(r),
\end{equation}
the action (\ref{actionfinal}) after angular integration reads
\begin{equation}\label{actionfinal2}
 S=\frac{1}{4 G}\int
dtdr
\frac{N(r)r^2}{\sqrt{h(r)}}\left[\mathcal{H}(r)+F(\mathcal{H}(r))\right].
\end{equation}
 Varying (\ref{actionfinal2})  with respect to $N(r)$ and setting
$\delta S/\delta N(r)=0$ we obtain
 \begin{eqnarray}
\mathcal{H}(r)+F(\mathcal{H}(r))=0. \label{varyN}
 \end{eqnarray}

\noindent
Before proceeding further, we should stress that we would have obtained the
same equation as above or other Einstein's equations, if we first construct
the field equation from the action given by equation (5), and then put the
ansatz (6).

Equation (9) is an algebraic equation for $\mathcal{H}(r)$ depending on
the functional form of $F$. One can always solve this equation to
get
 \begin{eqnarray}
\mathcal{H}(r)=\rm{constant},
\end{eqnarray}
which depends upon the functional form of $F(\mathcal{H}(r))$.
Denoting the above constant by the parameter $\beta$, we can
obtain
\begin{eqnarray}
&&\mathcal{H}(r)\equiv-\frac{2}{r^2}[X(r)r]^{\prime}=\beta\nonumber\\
&&\Rightarrow X(r)=-\frac{\beta r^2}{6}+\frac{A}{r},
 \end{eqnarray}
 where $A$ is
the integration constant, set from now on to $-2M$ with $M$ a new
constant. Thus,
\begin{eqnarray}
\label{frM}
 h(r)=1-\frac{2M}{r}-\frac{\beta r^2}{6}.
\end{eqnarray}

\noindent
Note that $\beta = 0$ case will give the standard Schwarzschild form for $h(r)$. Variation of (\ref{actionfinal2}) with respect to $h(r)$ leads to
\begin{eqnarray}
\label{ELeqn}
 \frac{d}{dr}\Big(\frac{\partial L}{\partial
h^{\prime}}\Big)-\frac{\partial L}{\partial h}=0,
\end{eqnarray}
where
\begin{eqnarray}
L=\frac{N(r)r^2}{\sqrt{h}}\Big[\mathcal{H}(h,h^{\prime},r)+F(\mathcal{H}(h,
h^{\prime},r)\Big].
\end{eqnarray}
A straightforward calculation gives
\begin{eqnarray}
\frac{\partial L}{\partial
h}=&-&\frac{1}{2}\frac{N(r)r^2}{h^{3/2}}\Big[\mathcal{H}+F(\mathcal{H})\Big
]\nonumber\\
&+&\frac{N(r)r^2}{\sqrt{h}}\Big[\frac{\partial
\mathcal{H}}{\partial R^{(3)}}+\frac{\partial
F(\mathcal{H})}{\partial R^{(3)}}\Big]\frac{\partial
R^{(3)}}{\partial h},
\label{2ndterm}
 \end{eqnarray}
  which under (\ref{varyN}) leads to
 \begin{eqnarray}
\frac{\partial L}{\partial
h}=\frac{2N(r)}{\sqrt{h}}\Big[1+\frac{\partial
F(\mathcal{H})}{\partial R^{(3)}}\Big].
\end{eqnarray}
Similarly, we acquire
\begin{eqnarray}
\frac{\partial L}{\partial h^{\prime}}&=&
\frac{N(r)r^2}{\sqrt{h}}\Big[\frac{\partial \mathcal{H}}{\partial R^{(3)}}+
\frac{\partial F(\mathcal{H})}{\partial R^{(3)}}\Big]\frac{\partial
R^{(3)}}{\partial h^{\prime}}\nonumber\\
&=&-\frac{2N(r)r}{\sqrt{h}}\Big[1+\frac{\partial
F(\mathcal{H})}{\partial R^{(3)}}\Big].
\label{1ndterm}
\end{eqnarray}
In conclusion, inserting (\ref{2ndterm}),(\ref{1ndterm}) into
(\ref{ELeqn}),
 and using that $1+\frac{\partial
F(\mathcal{H})}{\partial R^{(3)}}=\rm{constant}$ (arising from the
solution for $h(r)$ in $R^{(3)}$) we finally obtain
\begin{equation}
\frac{d}{dr}\Big(\frac{N(r)}{\sqrt{h}}\Big) = 0\ \ \Rightarrow\ \
N^{2}(r) = h(r).
\end{equation}
 This result,
together with equation  (\ref{frM}), shows that the static,
spherically-symmetric vacuum solution for a general $f(R)$ model
under ADM decomposition is either a Schwarzschild de-Sitter or
Schwarzschild anti-de-Sitter one, depending upon the choice of
$\beta$. Note that such Schwarzschild or Schwarzschild de-Sitter space
solution conditions for arbitrary models of $f(R)$ gravity had earlier also
been 
found out in \cite{cenoz}, although not in the context of
Ho\v{r}ava-Lifshitz modified gravity. We again stress that equation (10)
can have different algebraic solutions depending upon the form for F that
one chooses for the modified gravity part in action (5). But this will only
change the value of the constant $\beta$. It may happen for some choices of
$F$, $\beta$ may be imaginary and those solutions are unphysical. But for
all those cases, where $\beta$ is real, the solution of the metric will
always be of the form given by equation (12) and (18) irrespective of the
form for $F$.

Also note that we have used the ansatz to be static and spherically
symmetric in nature, although in standard cases, static is generally
understood by assuming spherical symmetry 
by the use of Birkhoff theorem. However, in standard metric modified
gravity models Birkhoff theorem does not hold good.

\section{Homogeneous and Isotropic Cosmological Evolution}
\label{cosmo}

 Here we assume that the background is homogeneous and
isotropic and the spatial 3-hypersurface is flat ($k=0$):
\begin{equation}
\label{metricFRW}
 ds^{2} = -N^{2}(t) dt^{2} + a^{2}(t)(dx^{2} +
dy^{2}+ dz^{2}),
 \end{equation}
 where $a(t)$ is the scale factor
and $N(t)$ is the lapse function. In order to present an example
of explicit cosmological solutions, we consider the form for
$f(R)$ proposed by Starobinsky in \cite{starob}:
\begin{equation}
\label{Starob}
 f(R) = \lambda R_{0}
\left\{\left[1+\left(\frac{R}{R_{0}}\right)^{2}\right]^{-n}-1\right\},
\end{equation}
 where $\lambda$, $R_{0}$ and $n$ are the model parameters, and
 from now on, $R$ denotes the four dimensional Ricci scalar.
 The advantage of choice (\ref{Starob}) is that it accepts a theoretical
justification, but one could also use a
 different ansatz at will.
 Using
this choice for $f(R)$, and under the metric (\ref{metricFRW}),
the action (\ref{actionfinal}) becomes
\begin{eqnarray}
S=\frac{1}{16\pi G}\int dt d^3x N\sqrt{g^{(3)}}\Big\{ \lambda
R_{0}\Big[\Big(1+\frac{36 H^{4}}{N^{4}
R_{0}^{2}}\Big)^{-n}-1\Big]\nonumber\\
 -\frac{6H^{2}}{N^{2}}\Big\} + S_{m},\ \ \
 \label{actionFRW}
\end{eqnarray}
where $H \equiv \frac{\dot{a}}{a}$ is the  Hubble parameter. As in the
static case, here also one gets the same equation of motion by putting the ansatz (19) and (20) first in the action (5) and then vary the action with respect to different parameters, or first vary the action (5) to get the equation of motion and then put the ansatz (19) and (20).

Finally, as usual, the energy density and pressure for the matter
field are respectively defined as
 \begin{eqnarray}
\rho_{m} &=& - \frac{1}{\sqrt{g^{(3)}}}\frac{\delta S_{m}}{\delta{N}}\\
g_{ij} p_{m} &=& - \frac{2}{N}{\sqrt{g^{(3)}}}\frac{\delta
S_{m}}{\delta{g^{ij}}}.
\end{eqnarray}

Variation of action (\ref{actionFRW}) with respect to $N$, and the
subsequent fixing $N=1$ (as it is usual in cosmological
applications of the ``foliation preserving'' framework), leads to
the ``effective'' Friedmann equation
\begin{widetext}
\begin{eqnarray}
H^{2} +\frac{ \lambda R_{0}}{6}\Big[\Big(1+\frac{36
H^{4}}{R_{0}^2}\Big)^{-n}-1\Big] + \frac{24 \lambda n
H^{4}}{R_{0}}\Big(1+\frac{36 H^{4}}{R_{0}^2}\Big)^{-(n+1)}=
\frac{8\pi G}{3} \rho_m. \label{Fr}
\end{eqnarray}
\end{widetext}
According to the usual approach, one could also vary  the action
(\ref{actionFRW}) with respect to the second variable $g_{ij}$.
This procedure, although straightforward, leads to a complicated
result, which forbids its physical use. Alternatively, we prefer
to use
 the matter conservation equation:
\begin{equation}
\label{mattconserv}
 \ \dot{\rho_{m}} + 3H(\rho_{m}+p_{m}) = 0,
\end{equation}
assuming for simplicity, and without loss of generality, the
matter to be dust ($p_{m} = 0$), which is definitely justified for
late-time cosmological behavior. We should mention here that, in our
original 
action (5), matter is minimally coupled to gravity via metric and it is
always
separately conserved. We stress that using
(\ref{mattconserv}) together with (\ref{Fr}), one can always
obtain the same equation with that arising from varying the action
(\ref{actionFRW}) w.r.t $g_{ij}$.

 In this case,
(\ref{mattconserv}) leads to the usual evolution $\rho_{m} =
\rho_{m0} a^{-3}$, with $\rho_{m0}$ the matter energy density at
present, where the scale factor is fixed to $1$. Therefore,
inserting this formula into (\ref{Fr}) we result to
\begin{widetext}
\begin{eqnarray}
\frac{H^{2}}{H_{0}^2} +  \frac{\lambda\alpha}{6}
\Big[\Big(1+\frac{36 H^{4}}{R_{0}^2}\Big)^{-n}-1\Big] + \frac{24
\lambda n}{\alpha}\Big(\frac{H}{H_{0}}\Big)^4 \Big(1+\frac{36
H^{4}}{R_{0}^2}\Big)^{-(n+1)} = \Omega_{m0}a^{-3},
 \label{Fr2}
\end{eqnarray}
\end{widetext}
 where  we have defined $\alpha \equiv \frac{R_{0}}{H_{0}^2}$ with $H_0$
the current $H$-value and $\Omega_{m0}$
the matter density-parameter at present. Thus, the parameter
$\alpha$ accounts for the modification of gravity. Finally, note
that imposing (\ref{Fr2}) at present, that is taking $a=1$ and
$H=H_0$, allows for the elimination of the parameter $\lambda$ in
favor of $\alpha$, $n$ and $\Omega_{m0}$, namely:
\begin{eqnarray}
\lambda=\frac{\Omega_{m0}-1}{\frac{\alpha}{6}\left[\left(1+\frac{36}{
\alpha^2}\right)^{-n}-1\right]+
\frac{24n}{\alpha}\left(1+\frac{36}{\alpha^2}\right)^{-(n+1)}}.
 \label{lambda}
\end{eqnarray}

Equation (\ref{Fr2}) is  the modified Friedmann equation of the
modified-gravity model at hand, and contains all the cosmological
information of the system. It presents the significant advantage
that it does not contain any higher-order time-derivatives that
appear in the usual approach of $f(R)$-gravity models. Instead,
and due to ADM decomposition, it is just an algebraic equation for
the Hubble-parameter $H(a)$, although of not simple form.
Therefore, one can examine its solutions for various values of
$n$, which in turn can be used to construct all the observable
quantities like deceleration parameter, luminosity distance,
angular diameter distance etc.

In particular, knowledge of  $H(a)$ allows for a straightforward
calculation of $dH(a)/da\equiv H'(a)$, while for every quantity
$Q$ we obtain $\dot{Q}=Q'(a) a H(a)$. Therefore, for the
deceleration parameter $q\equiv-\ddot{a}/[a H(a)]^2$ we acquire:
\begin{equation}
q(a)=-1-\frac{a}{H(a)}H'(a). \label{qa}
\end{equation}
As usual, $q<0$ corresponds to $\ddot{a}>0$ that is to an
accelerating universe, while $q>0$ corresponds to a decelerating
one. Finally, $q<-1$ corresponds to $\dot{H}>0$, which is the case
of a super-accelerating universe  \cite{Das:2005yj}.

As it is usual for modified gravity models, the role of matter is
crucial for the determination of the cosmological behavior. For
instance, in the complete absence of matter, that is setting
$\Omega_{m0}$ to zero, (\ref{Fr2}) implies immediately that $H$ is
independent of $a$, that is $H'(a)=0=\dot{H}$ and $q=-1$ which is
just what is expected in this case. In the following we focus on
the realistic case where $\Omega_{m0}\approx 0.3$.

As can be seen from  (\ref{Fr2}), even for $n=1$ the equation for
$H(a)$ is of high order, giving rise to many solution branches,
the number of which increases fast with increasing $n$. Some of
these solution-branches lead to imaginary $H(a)$ and thus are not
physical. Additionally, one gets solution-branches that lead to
divergences in finite scale factors in the past, which must also
be omitted. We are interested in those branches that have $H(a)>0$
at all $a$ and $q(a)<0$ at large $a$, that is corresponding to an
expanding universe which accelerates at late cosmological epochs.
One significant advantage of the aforementioned procedure is that
since we do not solve any differential equation, we do not have to
face the discussion of the initial-condition determination. We
only have to fix the values of $H_0$, $\Omega_{m0}$ at present,
and then the system is fully determined for particular values of
$n$
 and $\alpha$.

\begin{figure}[ht]
\begin{center}
\mbox{\epsfig{figure=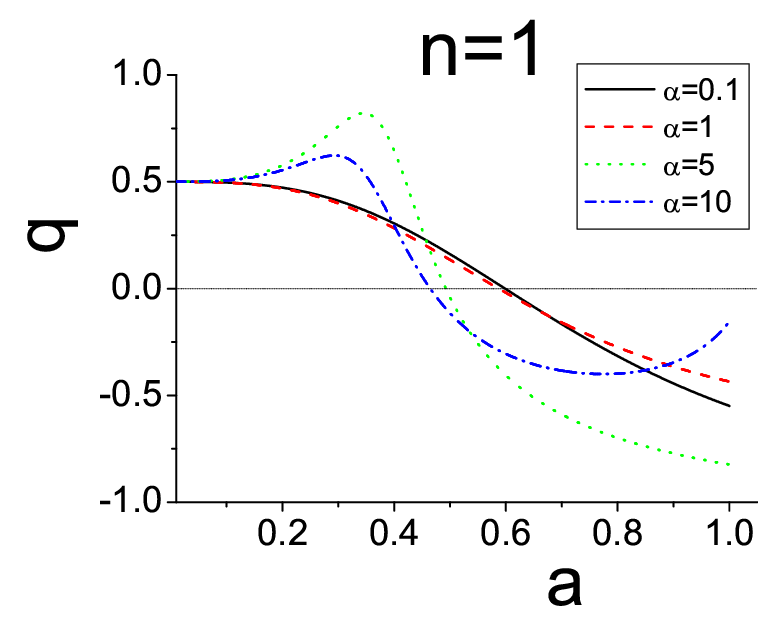,width=8.cm,angle=0}} \caption{{\it
 The deceleration parameter $q(a)$ for an expanding universe for  $n=1$,
with $\Omega_{m0}= 0.3$ and  $H_0=1$, $a_0=1$.
The curves correspond to $\alpha=0.1$ (black-solid), $\alpha=1$
(red-dashed), $\alpha=5$ (green-dotted) and $\alpha=10$
(blue-dashed-dotted). The horizontal line marks the $q=0$ bound.
}} \label{n1}
\end{center}
\end{figure}

Let us first investigate the $n=1$ solution subclass. In
fig.~\ref{n1} we depict the deceleration parameter $q(a)$ as it is
given by  (\ref{qa}) for the numerically obtained solution of the
algebraic equation (\ref{Fr2}) for $n=1$ with $\Omega_{m0}= 0.3$. The
curves correspond to four values of the parameter $\alpha$.
As we observe, in all cases we do obtain acceleration at late
times, with the transition to the accelerated phase realized at
earlier times for larger $\alpha$-values, that is for more
significant gravity-modification. It is interesting to notice that
for large values of $\alpha$, the behavior of $q(a)$ can be
non-monotonic (as can be seen in the $\alpha=10$ curve), and going
to even larger values ($\alpha=20$) it brings deceleration at very
late times. Since this scenario is not favored by observations we
do not show it explicitly, but it is characteristic of the rich
phenomenology and cosmological possibilities that our model
presents. Finally, we mention that in all cases $q(a)$ is larger
than $-1$, that is the imposed $f(R)$-ansatz does not seem to be
able to lead to a super-accelerated universe.

\begin{figure}[ht]
\begin{center}
\mbox{\epsfig{figure=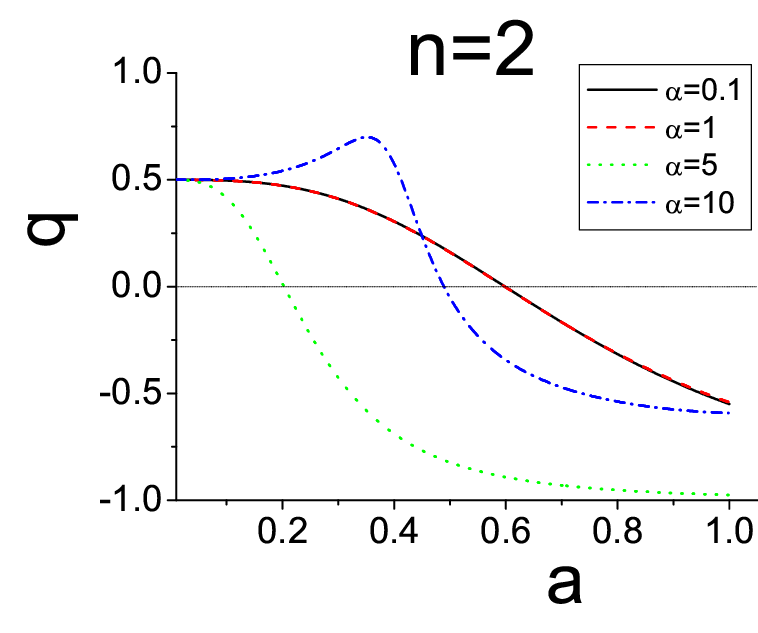,width=8.cm,angle=0}} \caption{{\it
 The deceleration parameter $q(a)$ for an expanding universe for  $n=2$,
with $\Omega_{m0}= 0.3$ and  $H_0=1$, $a_0=1$.
The curves correspond to $\alpha=0.1$ (black-solid), $\alpha=1$
(red-dashed), $\alpha=5$ (green-dotted) and $\alpha=10$
(blue-dashed-dotted). The horizontal line marks the $q=0$ bound.}}
\label{n2}
\end{center}
\end{figure}

In fig.~\ref{n2} we depict the $q$-behavior for the $n=2$ solution
subclass.
In this case, the curves for $\alpha=0.1$ and $\alpha=1$ have a
small difference (not observed in the scale of the figure), and
one needs to go to larger $\alpha$ in order to see a different
cosmological evolution (which is achieved fast for $\alpha$
becoming larger that $1$). As we see, we again obtain acceleration
at late cosmological epochs. Note that for $\alpha=5$, the
transition to the acceleration phase is realized earlier than for
$\alpha=10$, as a result of the highly non-linear behavior of
equation (\ref{Fr2}).

\begin{figure}[ht]
\begin{center}
\mbox{\epsfig{figure=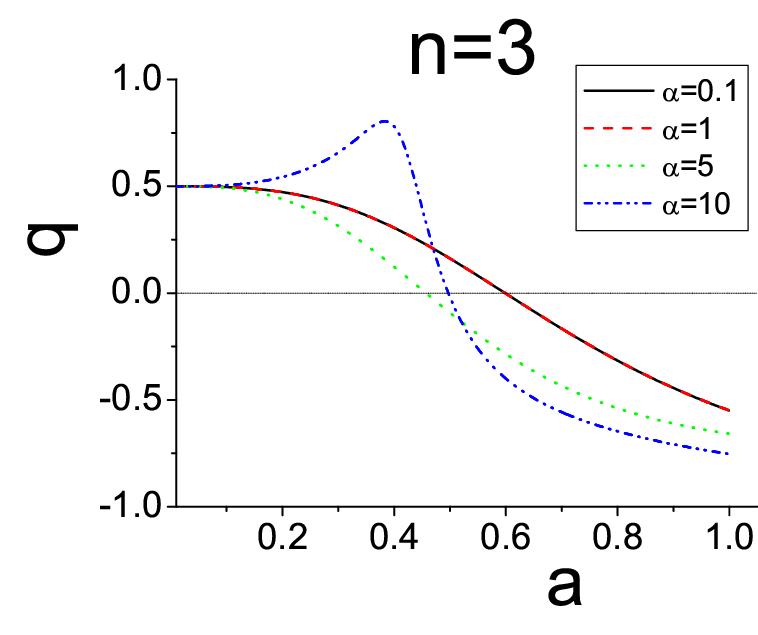,width=8.cm,angle=0}} \caption{{\it
 The deceleration parameter $q(a)$ for an expanding universe for  $n=3$,
with $\Omega_{m0}= 0.3$ and  $H_0=1$, $a_0=1$.
The curves correspond to $\alpha=0.1$ (black-solid), $\alpha=1$
(red-dashed), $\alpha=5$ (green-dotted) and $\alpha=10$
(blue-dashed-dotted). The horizontal line marks the $q=0$ bound.}}
\label{n3}
\end{center}
\end{figure}

In fig.~\ref{n3} we depict the $q$-behavior for the $n=3$ solution
subclass.
Similarly to the previous cases, we do obtain late-time
acceleration, with $q(a)$ for large $\alpha$ being non-monotonic.
Finally, we mention that qualitatively similar results arise for
larger values of $n$ too, but for simplicity we do not present
them explicitly.

The aforementioned solutions correspond to $H(a)>0$ at all $a$'s,
that is to an expanding universe. However, for completeness we
mention that the present model allows also for solutions that
describe a contracting universe. Indeed, since (\ref{Fr2}) is a
even equation for $H(a)$, we deduce that for every
$H(a)$-solution, $-H(a)$ is also a solution. Therefore, while the
investigated solutions of figures \ref{n1} to \ref{n3} possess
$H(a)>0$ at all scale factors, the same $q(a)$-behavior arise from
the corresponding branches with $H(a)<0$ at all $a$, that is a
contracting universe.

\section{Conclusions}
\label{Conclusions}

In the present work we have studied  $f(R)$-gravity models
performing the ADM slicing of standard General Relativity, that is
its (3+1)-decomposition based on the Hamiltonian formulation. This
approach allows for an easier treatment of modified-gravity
systems and for the extraction of their general theoretical and
cosmological implications.

As a first application we derived the static,
spherically-symmetric vacuum solutions for general
$f(R)$-ansatzes. As we saw, they correspond to either
Schwarzschild de-Sitter or Schwarzschild anti-de-Sitter ones,
depending upon the choice of a particular parameter $\beta$.

Concerning applications in a cosmological framework, we
investigated the evolution of a homogeneous and isotropic flat
universe. Imposing as a specific example a particular ansatz for
$f(R)$, we showed that the Hubble parameter is given by an
algebraic equation in terms of the scale factor, with a new
parameter that determines the modification of gravity. This fact
is an advantage since one does not need to discuss the initial
condition, since all the information is included in the value of
the matter density parameter  at present. The system accepts many
solution branches, the physical sub-class of which corresponds to
either expanding or contracting universes. Furthermore, one can
easily acquire solutions that correspond to acceleration at late
cosmological epochs, in agreement with observations, and this is
achieved without the need of fine tuning the model-parameters or
the initial conditions. The model at hand presents rich
cosmological behavior, and moreover one can calculate additional
observables such as luminosity distance and angular diameter
distance.

In conclusion, motivated by the Horava gravity \cite{horava1}, we have 
studied a new approach for the $f(R)$ modified gravity, where the modified
action satisfies the foliation preserving diffeomorphism instead of the
full 4D general diffeomorphism. This is in accordance to what proposed
by Ho\v{r}ava to improve the UV behavior of gravity. With such a
modification we have showed that the field equations are still second
order, unlike the standard $f(R)$ gravity, and hence the present scenario
does not contain any extra scalar degree of freedom. This makes the model
compatible with the local gravity experiments, but still having interesting
cosmological consequences. 
\\

\paragraph*{{\bf{Acknowledgements:}}}
A.A.S acknowledges the  financial support provided by the
University Grants Commission, Govt. Of India, through major
research project grant (Grant No:33-28/2007(SR)). SKC would like
to thank ICTP, Trieste for warm hospitality during a visit where a
part of this work was done.

\end{document}